\documentclass[reprint]{revtex4-1}
\usepackage{amsmath,amssymb,graphicx}
\usepackage{graphicx}
\usepackage{dcolumn}
\usepackage{bm}
\usepackage{booktabs}

\begin{document}

\preprint{APS/123-QED}

\title{Observation of magneto-optical rotation effects in cold $^{87}$Rb atoms in an integrating sphere}
\author{Jinyin Wan, Yanling Meng, Peng Liu, Xiumei Wang, Yaning Wang, Ling Xiao}
\author{Huadong Cheng}
\email{corresponding author: chenghd@siom.ac.cn}
\author{Liang Liu}
\email{corresponding author: liang.liu@siom.ac.cn}
 \affiliation{Key Laboratory of Quantum Optics and Center of Cold Atom Physics, Shanghai Institute of Optics and Fine Mechanics, Chinese Academy of Sciences, Shanghai 201800, China.}

\begin{abstract}
We present a modified scheme for detection of the magneto-optical rotation (MOR) effect, where a linearly polarized laser field is interacting with cold $^{87}$Rb atoms in an integrating sphere. The rotation angle of the probe beam's polarization plane is detected in the experiment. The results indicate that the biased magnetic field, the probe light intensity and detuning, and the cold atoms' temperature are key parameters for the MOR effect. This scheme may improve the contrast of the rotation signal and provide an useful approach for high contrast cold atom clocks and magnetometers.

\end{abstract}

\maketitle


\section{Introduction}
Recently, MOR effect has played an important role in the study of light scattering in the direction of the incident light~\cite{Budker02}. It is an interaction between light and magnetic field in a medium, the faraday effect causes a rotation of polarization plane which is linearly proportional to the component of the magnetic field in the direction of propagation. It shows higher sensitivity than the absorption measurement~\cite{Church74}. The improved sensitivity is mainly due to the almost complete elimination of the background light. Lin et al. have used faraday rotation method to detect clock signal in a vapor cell to improve its contrast to 90\%~\cite{Lin12}. We have observed faraday rotation signal in cold $^{87}$Rb atoms in an integrating sphere and obtained a Ramsey fringe whose contrast is 92\%~\cite{Zheng14}. The magneto-optical signal with cold atoms can also be interpreted by Larmor's theorem as a sensitive rotation of the atomic wavefunction reflecting the magnetic field, Optical Magnetometery techniques with Rubidium Bose-Einstein condensates (BEC) have recently been demonstrated with high spatial resolution~\cite{Wildermuth06,Vengalattore07}. Then such detection scheme may provide a feasible approach for high contrast cold atom clocks and magnetometers.

In this work, we optimized and modified the previous experiment~\cite{Zheng14}. The main improvement is that: Firstly, an independent probe laser is used to eliminate additional noise, since both of the cooling and probe lights are from the same extended-cavity diode laser in Ref.~\cite{Zheng14}; Secondly, the rotation angle is modified as the transmitted probe light divided by the corresponding background probe light intensity under different conditions. While in Ref.~\cite{Zheng14}, it is roughly approximated as the transmitted probe light's power divided by the incident probe light's power, there is no pumping light to get the precise background probe light intensity. It is not rigorous cause different probe detunings controlled by AOM produce different background probe light intensities; Thirdly, we compare the experimental results with the theoretical calculations and find they agree well with each other.
\section{Theory}

Generally, laser cooling of atoms directly from a vapor background in diffuse laser light does not require precise collimation of laser beams and therefore has received a lot of attention for its application of compact cold atom clocks~\cite{Guillot01,Cheng09}. The forward scattering (FS) intensity of the weak probe light can be written as~\cite{Budker02}
\begin{equation}\label{I}
\begin{split}
 I_{FS}=&\frac{1}{4}I_{0}(e^{-\alpha_{+}\omega l/c}-e^{-\alpha_{-}\omega l/c})^{2}+\\
 &I_{0}e^{(-\alpha_{+}+\alpha_{-})\omega l/c}\sin^{2}[\frac{(n_{+}-n_{-})\omega l}{2c}],
\end{split}
\end{equation}
where $I_{0}$ is the probe light intensity, $\omega$ is the angular frequency of the probe light, $l$ is the length of the cold atom vapor, $\alpha_{\pm}$ and $n_{\pm}$ are the absorption and refractive indices for the $\sigma^{\pm}$ components of the probe light, respectively. The polarizers are assumed to be ideal here. We approximate the cold atom as a two-level closed non-degenerate system, the absorption and refractive indices can be written as ~\cite{Scully97}
\begin{subequations}
\begin{align}\label{n}
\alpha_{+}&=\frac{P^{2}\Delta N}{\varepsilon_{0}\hbar} \frac{\gamma}{\gamma^{2}+(\Delta \omega+2\delta)^{2}},\\
\alpha_{-}&=\frac{P^{2}\Delta N}{\varepsilon_{0}\hbar} \frac{\gamma}{\gamma^{2}+(\Delta \omega)^{2}},\\
n_{+}&=\frac{P^{2}\Delta N}{\varepsilon_{0}\hbar} \frac{\Delta \omega+2\delta}{\gamma^{2}+(\Delta \omega+2\delta)^{2}},\\
n_{-}&=\frac{P^{2}\Delta N}{\varepsilon_{0}\hbar} \frac{\Delta \omega}{\gamma^{2}+(\Delta \omega)^{2}},
\end{align}
\end{subequations}

where $P$ is the dipole moment, $\varepsilon_{0}$ is the free space permittivity, $\gamma$($\sim$~0.3~s$^{-1}$) is the relaxation rate of the ground states, $\Delta \omega$ is the detuning of the probe light, $\delta$(=$g\mu_{B}B/\hbar$) is the detuning from $\sigma^{\pm}$ caused by Zeeman shifts, where $g$ is the lande g-factor, $\mu_{B}$ is the Bohr magneton, $B$ is the biased magnetic field. $\Delta N$(=$N_{0}\rho(\Omega_{l})$, $\Omega_{l}\ll\Gamma^{*}$) is the population difference between the ground and excited states, here $N_{0}$ is the atomic concentration of the cold atoms, $\rho(\Omega_{l})=\Omega_{l}^{2}/(4\gamma'^{3}\Gamma^{*})$ is the density matrix element which represents the population in ground state $5^{2}S_{1/2}, |F=2\rangle$ at dynamic equilibrium and can be obtained by solving the optical Bloch equation~\cite{Arimondo96}, where $\Omega_{l}$ is the Rabi frequency of the probe light, $\Gamma^{*}=3.8\times10^{7} s^{-1}$, and $\gamma'=\gamma+(|\Omega_{l}|^{2}/2\Gamma^{*})\sim\gamma$. The exponential part in Eq.~\eqref{I} corresponds to the absorption, and the sine function describes the birefringence. Since the rotation angle is very small, it could be approximated as
\begin{equation}\label{r}
\theta\approx\frac{I_{FS}}{I_{0}}.
\end{equation}
As can be seen from Eqs.~\eqref{I}-\eqref{r}, the rotation signal mainly depends on the following parameters: the applied biased magnetic field, the probe light intensity, and the detuning of probe light. The rotation angles' relationships with the above parameters are studied in the following sections.

\section{Experimental Details}

\begin{figure}[htbp]
\centerline{\includegraphics[width=0.48\textwidth]{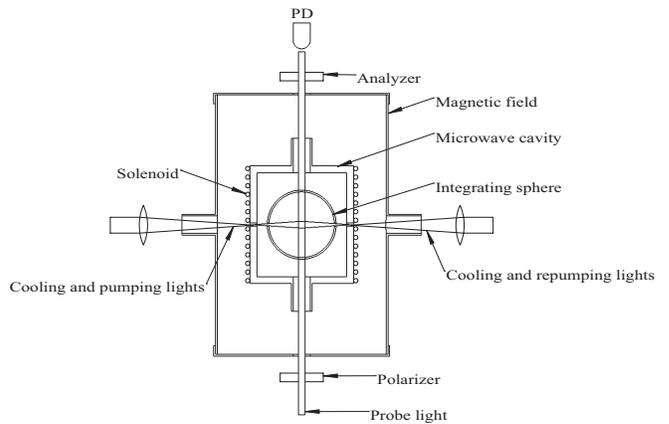}}
\caption{Schematic of the experimental setup (not scaled). The integrating sphere and solenoid are placed inside a cylindrical $\mu$-metal magnetic shield to isolate them from external magnetic fields. The solenoid wrapping around the sphere produces the longitudinal biased magnetic field. Before and after the physical system, a linear polarizer (Glan-Taylor prism) and an analyzer which have orthogonal optical axes, acting as a blocker to the probe light background from the PD.}\label{Fig-experimental}
\end{figure}

The detection scheme is briefly described as: A linearly polarized laser passes through and interacts with cold $^{87}$Rb atoms in an integrating sphere, and the transmitted beam is detected by a photodetector (PD). From the detector's output, The rotation angle of the probe beam's polarization plane is extracted. Fig.~\ref{Fig-experimental} shows the schematic diagram of the basic experimental setup. Cold $^{87}$Rb atoms are prepared in an integrating sphere whose surface is sprayed by the reflective material Avian B from Avian Technologies with diffuse reflective index of $>$98\% at 780~nm. The sphere's inner diameter is $40$~mm and the background pressure is maintained at $\sim10^{-7}$~Pa. The cooling and repumping beams are combined by a polarization beam splitter (PBS), then they are evenly split into two parts and injected into the integrating sphere through two multimode fibers. The linearly polarized probe beam passes through the sphere which lies between two crossed polarizers. The solenoid wrapping around the sphere produces longitudinal magnetic field that provides a quantization axis and lifts the Zeeman sub-level degeneracy. The transmitted FS signal depends on the magneto-optical rotation of the polarization plane of the incident linearly polarized probe light whose background noise is almost eliminated~\cite{Corney66,Gawlik74}, which is a result of the birefringence caused by the opposite displacement of the dispersion curves of the left and right circular polarizations~\cite{Budker02,Wieman76}.

\begin{figure*}[htbp]
\centerline{\includegraphics[width=0.95\textwidth]{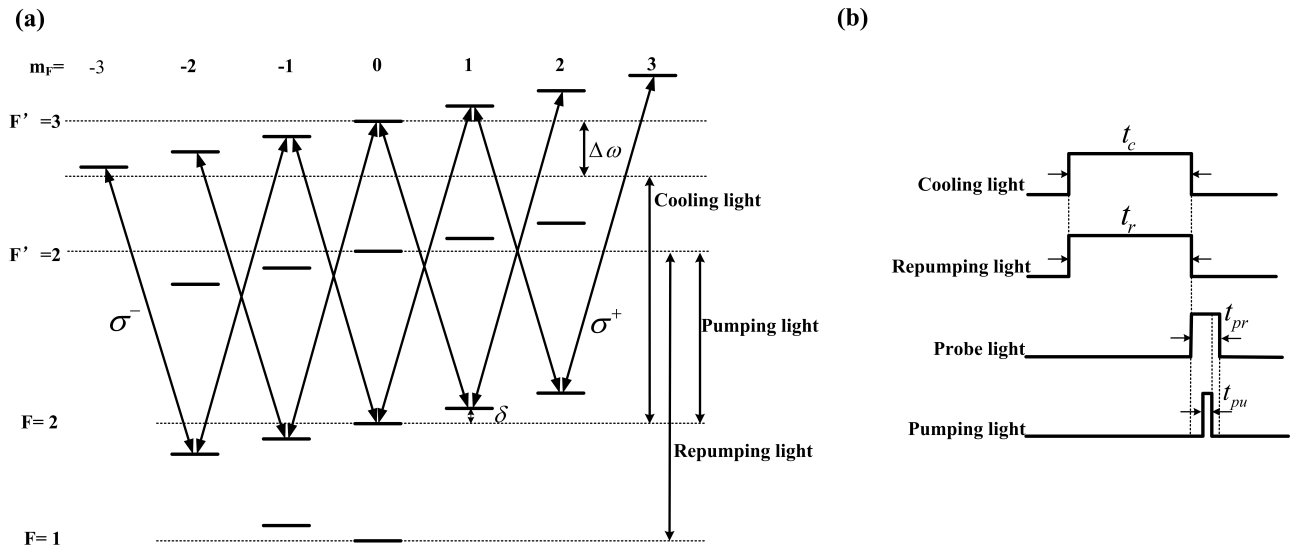}}
\caption{(a) Energy levels of $^{87}$Rb $D_{2}$ line interacting with lasers and the biased magnetic fields (not scaled). (b) Timing sequence of the experiment (not scaled). $t_{c}=t_{r}=$ 177.5~ms, $t_{pr}=$ 7.5~ms, and $t_{pu}=$ 2~ms.}\label{Fig-energy}
\end{figure*}

\begin{figure}[htbp]
\centerline{\includegraphics[width=0.48\textwidth]{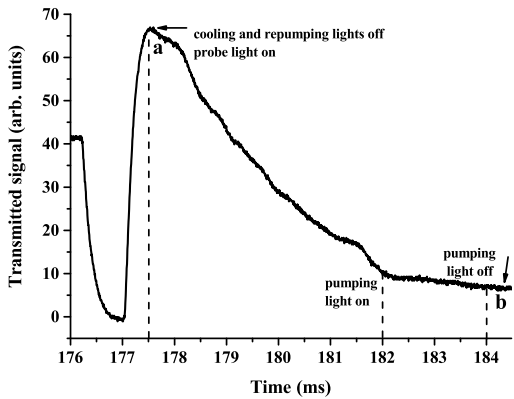}}
\caption{Transmitted signal versus time with what time lasers are switched on and off indicated. The magnetic field $B$ is 27.4~mG, the probe light detuning and intensity are -15~MHz and 117~$\mu$W/cm$^{2}$, respectively. The absorption signal is calculated as point a minus point b (a-b).}\label{Fig-absorption}
\end{figure}

The cooling and pumping beams are provided by an extended-cavity diode laser (Topica TA100), their detunings and intensities are time-sequence controlled by two different acousto-optic modulators (AOMs), respectively. The probe (Toptica DL pro) beam's detuning and intensity are time-sequence controlled by an AOM, and the repumping (Topica DL100) beam are time-sequence controlled by a mechanical shutter. $^{87}$Rb atomic levels are illustrated in Fig.~\ref{Fig-energy}(a). In the presence of the weak longitudinal magnetic field, circular polarization lights ($\sigma^{\pm}$) excite the transitions among states $5^{2}S_{1/2}, |F=2, m_{F}=0,~\pm1,~\pm2\rangle$ and $5^{2}P_{3/2}, |F'=3, m_{F}=0,~\pm1,~\pm2,~\pm3\rangle$. The cold Rubidium atoms are prepared by the cooling light locked to the red-detuning of transition $5^{2}S_{1/2}, |F=2\rangle\rightarrow5^{2}P_{3/2}, |F'=3\rangle$ and the repumping light locked to the transition $5^{2}S_{1/2}, |F=1\rangle\rightarrow5^{2}P_{3/2}, |F'=2\rangle$. The probe beam is is also red-detuned to the transition of $5^{2}S_{1/2}, |F=2\rangle\rightarrow5^{2}P_{3/2}, |F'=3\rangle$ and its diameter is expanded to 4~mm. The pumping light is locked to the transition of $5^{2}S_{1/2}, |F=2\rangle\rightarrow5^{2}P_{3/2}, |F'=2\rangle$ to keep atoms in the ground state $5^{2}S_{1/2}, |F=1\rangle$. The powers of cooling, repumping and pumping lights are 130, 5 and 1~mW, respectively. The timing sequence of the experiment is shown in Fig.~\ref{Fig-energy}(b), cooling time $t_{c}$ and repumping time $t_{r}$ is 177.5~ms, and probe time $t_{pr}$ is 7.5~ms, the probe beam is switched on once after the cooling and repumping lights are turned off. Pumping time $t_{pu}$ is 2~ms, the pumping light is switched on after the probe laser lasts 4.5~ms to precisely detect the background probe light intensity. While in Ref.~\cite{Zheng14}, there is no pumping light in the detecting process. The transmitted signal versus time in a cycle time is shown in Fig.~\ref{Fig-absorption} when the magnetic field $B$ is 27.4~mG, the probe light detuning and intensity are -15~MHz and 117~$\mu$W/cm$^{2}$, respectively. The absorption is calculated as point a minus point b (a-b). The real background position b is obtained after the pumping time, while if there is no pumping light, one could not get the precise absorption signal. In this work, the experimental rotation angle is calculated as $I_{FS}$ calculated from the absorption signal divided by the corresponding $I_{0}$ which is measured from the PD when only probe light is on.

\section{Experimental Results and Discussions}
\begin{figure}[htbp]
\centerline{\includegraphics[width=0.48\textwidth]{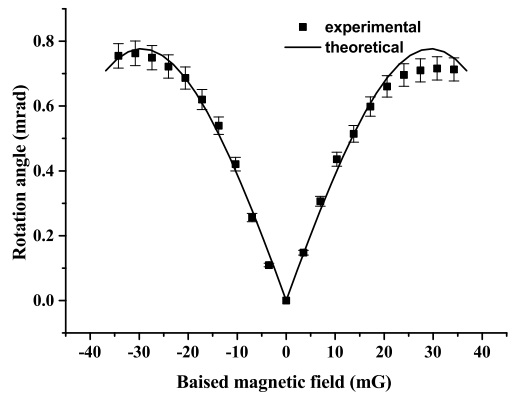}}
\caption{Experimental (squares, the uncertainty of the data points is 5\%) and theoretical (solid curve) results of rotation angles versus the biased magnetic fields. The rotation angle $\theta$ is about symmetric with the two opposite magnetic field directions, it is linear with $B$ at small biased magnetic field, peaks at 30.8~mG and then falls off.}\label{Fig-magnetic}
\end{figure}

In the experiment, we detected the relationships of the MOR effect shown as the detected rotation angle, and investigated its relationships with the applied biased magnetic field, the probe light intensity, and the probe light detuning according to Eqs.~\eqref{I}-\eqref{r}. Fig.~\ref{Fig-magnetic} shows rotation angles versus the biased magnetic field when the probe light's detuning is -15~MHz and its intensity is about 170~$\mu$W/cm$^{2}$. The theoretical results (solid curve) according to Eq.~\eqref{r} fit well with the experimental results (squares), here the calculated curve depends on the energy levels illustrated in Fig.~\ref{Fig-energy}(a). The rotation angle $\theta$ is about symmetric as for the two opposite magnetic field directions. It is linear with the amplitude of $B$ at small values and the slope is about 0.0325 mrad/mG which represents the sensitivity of faraday rotation angle versus the biased magnetic field, the rotation angle peaks at about 30.8~mG and then falls off at large magnetic fields. Such results in cold atoms are similar to that in vapor cell schemes~\cite{Budker02}, it shows that an optimal magnetic fields exists (30.8~mG in our experimental conditions) and larger magnetic fields would decrease the rotation signal, which is an universal phenomenon either in vapor cell or cold atom samples. While in Ref.~\cite{Zheng14}, it sees a linear trend of the rotation angle versus the biased magnetic field which gets a somewhat misleading result.
\begin{figure}[htbp]
\centerline{\includegraphics[width=0.49\textwidth]{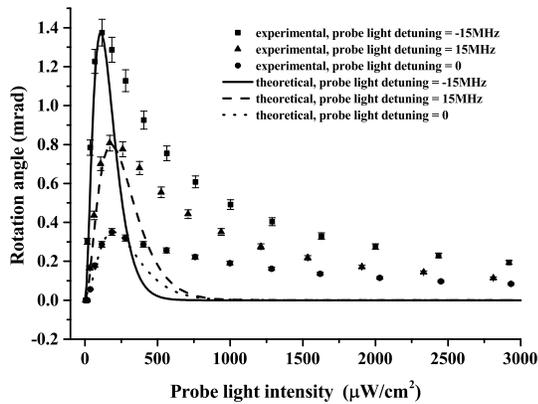}}
\caption{Experimental (symbols, the uncertainty of the data points is 5\%) and theoretical (curves) results of rotation angle versus probe light's intensity when $B=27.4$~mG and the probe light detuning is -15~MHz, 15~MHz, and zero (on resonance), respectively. Each curve has a peak at about the same probe light intensity of 117~$\mu$W/cm$^{2}$ and then decreases. }\label{Fig-intensity}
\end{figure}
Fig.~\ref{Fig-intensity} shows the experimental (symbols) and the corresponding theoretical (curves) results of rotation angles versus the probe light's intensities when $B=27.4$~mG and the probe light detuning is -15, 15 and 0~MHz (on resonance), respectively. The experimental results agree well with theoretical calculations. Each curve has a peak at the same probe light intensity of $\thicksim$117~$\mu$W/cm$^{2}$ and then decreases quickly to different fixed values at large probe light intensities in a descending trend. Nonlinear effects appear since the rotation angle becomes intensity dependent.

\begin{figure}[htbp]
\centerline{\includegraphics[width=0.48\textwidth]{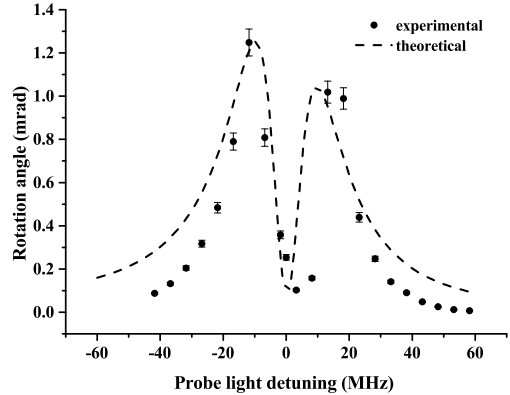}}
\caption{Experimental (circles, the uncertainty of the data points is 5\%) and theoretical (dashed curves) and results of the rotation angle versus the probe light's detunings when the probe light intensity is 117~$\mu$W/cm$^{2}$ and the biased magnetic field is 27.4~mG. There are two peaks and a dip at detunings of -15~MHz, 10~MHz, and 0, respectively.}\label{Fig-detuning1}
\end{figure}

\begin{figure}[htbp]
\centerline{\includegraphics[width=0.48\textwidth]{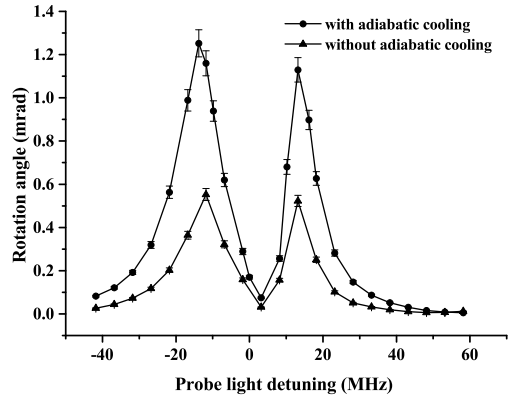}}
\caption{The rotation angles versus the probe light detunings with (circles) and without (triangles) adiabatic cooling when B=~-27.4~mG and $I_{0}=117~\mu$W/cm$^{2}$.}\label{Fig-detuning2}
\end{figure}

Fig.~\ref{Fig-detuning1} shows the experimental (circles) and theoretical (dashed curve) results of rotation angles versus the probe light detunings when the biased magnetic field is 27.4~mG and the probe light intensity is 117~$\mu$W/cm$^{2}$. The experimental profile fits well with the theoretical curve. There are two peaks and a minimum appear at the probe light detunings of -15, 0 and 0~MHz, respectively. It is consistent with curves in Fig.~\ref{Fig-intensity} that the largest rotation angle lies at the probe light detuning of -15~MHz and the smallest is on resonance. Thus we modified the misleading results that the maximum peak is at probe light's blue detuning shown in Ref.~\cite{Zheng14}. Because in that work the rotation angle is roughly approximated as the transmitted probe light's power divided by the incident probe light's power, there is no pumping light to get the precise background probe light intensity. It is not rigorous cause different probe detunings controlled by AOM produce different background probe light intensities. And our modification is proved to be reasonable by the theoretical calculations. We also detect the rotation angles versus the probe light detunings with the same magnetic field magnitude but in the opposite directions as shown in Fig.~\ref{Fig-detuning2} (squares). Comparing with Fig.~\ref{Fig-detuning1}, the profiles are the same, it indicates that opposite magnetic field directions don't change the detection signals' trend.
Generally, temperature is an important parameter in the experiment of cold atoms. Ref.~\cite{Zheng13} has demonstrated that the temperature of cold atoms is lower with adiabatic cooling. So we detect the the rotation angle versus the probe light detuning with (circles) and without (triangles) adiabatic cooling as shown in Fig.~\ref{Fig-detuning2} when B=~-27.4~mG and $I_{0}=117~\mu$W/cm$^{2}$. it proves that the rotation signals with adiabatic cooling are much larger at the same probe light detunings, which indicates that lower temperature of cold atoms could get larger MOR effect.

\section{Conclusion}
In conclusion, we present a modified scheme detecting the MOR effect shown as the rotation angels in a cold $^{87}$Rb atom vapor in an integrating sphere experimentally and theoretically, where a linearly polarized laser field is interacting with cold $^{87}$Rb atoms. The rotation angles' relationships with the biased magnetic field, the probe light intensity, the probe light detuning, and the temperature of cold atoms are investigated, respectively. The results show that there have optimal parameters for the largest MOR effect at lower atom temperature with adiabatic cooling, the biased magnetic field is about 27.4~mG, the probe light intensity is 117~$\mu$W/cm$^{2}$, and the probe light detuning is -15~MHz, respectively. The experimental results are compared with the theoretical calculations and they agree well with each other, which rectifies some misleading results in Ref.~\cite{Zheng14}. Since the MOR detection scheme has proven to get better Ramsey fringes than that with the absorption method in the integrating sphere~\cite{Zheng14}, and the MOR signal with cold atoms indicates a sensitive rotation of the atomic wavefunction reflecting the magnetic field, such detection scheme may provide a promising method for high contrast cold atom clocks and magnetometers. We would investigate its applications in these two aspects in the next step.

\section*{Acknowledgements}
This work was supported by National High Technology Research and Development Program of China (Grant No. 2012AA120702).

\end{document}